\documentclass[aps,prd,amsmath,amssymb,superscriptaddress,altaffillsymbol, preprintnumbers,preprint,nofootinbib,a4paper]{revtex4}
\usepackage{amsthm}
\usepackage{graphicx}
\usepackage{color}
\newcommand{\beq}{\begin{eqnarray}}
\newcommand{\eeq}{\end{eqnarray}}

\newcommand{\bmp}{\noindent\begin{minipage}{16cm}}
\newcommand{\emp}{\end{minipage}\vskip 7mm} 

\usepackage{dcolumn}
\usepackage{bm}
\usepackage{slashed}

\theoremstyle{definition}

\theoremstyle{plain}

\usepackage{epsfig}

\usepackage[ margin=5pt, font=normalsize,labelfont=bf,justification=raggedright]{caption}

\usepackage{hyperref}
\definecolor{rossoCP3}{cmyk}{0,.88,.77,.40}
\definecolor{verdeCP3}{rgb}{0.09765625, 0.57421875, 0.1015625}
\definecolor{bluCP3}{rgb}{0, 0.23, 0.67}
\hypersetup{colorlinks, bookmarksopen, bookmarksnumbered,citecolor=verdeCP3, linkcolor=bluCP3, pdfstartview=FitH, urlcolor=rossoCP3}
\def\lsim{\mathrel{\rlap{\lower4pt\hbox{\hskip1pt$\sim$}}
    \raise1pt\hbox{$<$}}}                
\def\gsim{\mathrel{\rlap{\lower4pt\hbox{\hskip1pt$\sim$}}
    \raise1pt\hbox{$>$}}}                

\newcommand{\bea}{\begin{eqnarray}}
\newcommand{\eea}{\end{eqnarray}}

\newcommand{\ba}{\begin{eqnarray}}
\newcommand{\ea}{\end{eqnarray}}
    
\newcommand{\Tr}{\mbox{Tr}\;}
        
\newcommand{\be}{\begin{eqnarray}}
\newcommand{\ee}{\end{eqnarray}}

\begin{document}
\title{\Large  \color{rossoCP3} ~~\\ Fundamental Composite (Goldstone) Higgs Dynamics}
\author{Giacomo Cacciapaglia}
\email{g.cacciapaglia@ipnl.in2p3.fr}
\affiliation{\mbox{Universit\'e de Lyon, F-69622 Lyon, France: Universit\'e Lyon 1, Villeurbanne}\\ \mbox{CNRS/IN2P3, UMR5822, Institut de Physique Nucl\'eaire de Lyon.}} 
\author{Francesco  Sannino}
\email{sannino@cp3-origins.net} 
\affiliation{{\color{rossoCP3} CP$^{\, 3}$-Origins} \& Danish Institute for Advanced Study {\color{rossoCP3} DIAS}, University of Southern Denmark, Campusvej 55, DK-5230 Odense M, Denmark}

\begin{abstract}
We provide a unified description, both at the effective and fundamental Lagrangian level, of models of composite Higgs dynamics where the Higgs itself can emerge, depending on the way the electroweak symmetry is embedded, either as a pseudo-Goldstone boson or as a massive excitation of the condensate. We show that, in general, these states mix with repercussions on the electroweak physics and phenomenology. Our results will help clarify the main differences, similarities, benefits and shortcomings of the different ways one can naturally realize a composite nature of the electroweak sector of the Standard Model. We will analyze the minimal underlying realization in terms of fundamental strongly coupled gauge theories supporting the flavor symmetry breaking pattern SU(4)/Sp(4) $\sim$ SO(6)/SO(5). The most minimal fundamental description consists of an SU(2) gauge theory with two Dirac fermions transforming according to the fundamental representation of the gauge group. This minimal choice enables us to use recent first principle lattice results to make the first predictions for the massive spectrum for models of composite (Goldstone) Higgs dynamics. These results  are of the utmost relevance to guide searches of new physics at the Large Hadron Collider.
 \\[.1cm]
{\footnotesize  \it Preprint: CP$^3$-Origins-2014-003 DNRF90, DIAS-2014-3, LYCEN-2014-02}
 \end{abstract}

\maketitle

\section{Unified fundamental composite Higgs dynamics}


It is a fact that the Standard Model (SM) of particle interactions continues to be a very successful description of Nature. The discovery of the Higgs particle can be viewed as the crown jewel of its success. Nevertheless, if not from the experimental point of view, at least theoretically the Higgs sector of the SM is, to put it mildly, unappealing. Spontaneous symmetry breaking is not explained but simply modeled. Furthermore there is no field-theoretical consistent way to shield the electroweak scale from higher scales. This is the SM naturalness problem. For a mathematical classification of different degrees of naturality we refer to \cite{Antipin:2013exa}. 

A time-honored avenue to render the SM Higgs sector natural is to replace it with a fundamental gauge dynamics featuring fermionic matter fields.  The new dynamics is free from the naturalness problem. The oldest incarnation of this idea goes under the name of Technicolor \cite{Weinberg:1975gm,Susskind:1978ms}. In these models the Higgs-sector, and therefore the Higgs boson itself, are made by new fundamental dynamics. Variations on the theme appeared later in the literature \cite{Kaplan:1983fs,Kaplan:1983sm}. 

In the traditional Technicolor setup, the electroweak symmetry breaks thanks to the gauge dynamics of the new underlying gauge theory. The Technicolor Higgs is then identified with the lightest scalar excitation of the fermion condensate responsible for electroweak symmetry breaking. However the Technicolor theory per se is not able to provide mass to the SM fermions and therefore a new sector must be introduced. This new sector is important and can modify the physical mass of the Technicolor Higgs, typically reducing it \cite{Foadi:2012bb}. Another possibility is that the gauge dynamics underlying the Higgs sector does not break the electroweak symmetry but breaks a global symmetry of the new fermions: a Higgs-like state is therefore identified with one of the Goldstone Bosons (GB) of the global symmetry breaking. In this case the challenges are not only to provide masses to the SM fermions but also to break the electroweak symmetry in the first place via another sector which, in turn, should also contribute to give mass to the would--be pseudo-GB Higgs. In any event a true improvement with respect to the SM Higgs sector shortcomings would arise only if a more fundamental description exists. 

In this work we provide a first unified description of models of composite dynamics for the electroweak sector of the SM. This description will clarify the main differences, similarities, interplay, and shortcomings of the approaches. We will also provide specific underlying realizations in terms of fundamental strongly coupled gauge theories. This will permit us to use recent first principle lattice results to make the first predictions for the massive spectrum, which is of the utmost relevance to guide searches of new physics at the Large Hadron Collider (LHC). Furthermore we will also demonstrate that, for a generic vacuum alignment, the observed Higgs is neither a purely pGB state nor the technicolor Higgs, but rather a mixed state. This fact impacts on its physical properties and associated phenomenology.

As possible underlying gauge theories we will consider those featuring fermionic matter. The pattern of chiral symmetry breaking depends on the underlying gauge dynamics \cite{Peskin:1980gc,Preskill:1980mz,Kosower:1984aw}. Of course, the initial hypothesis that the global symmetry breaks dynamically should be verified. In fact we know that, depending on the number of matter fields, the choice of the underlying gauge group and the dimension of the gauge group (e.g. the number of underlying fermionic matter), the symmetry might not break at all because the theory can develop large distance conformality, as discussed in \cite{Sannino:2004qp} for fermions in two-index representations, in \cite{Dietrich:2006cm} for a universal classification for SU($N$) gauge theories and their applications to Technicolor and composite dark matter models, in \cite{Sannino:2009aw} for orthogonal and symplectic groups, and in \cite{Mojaza:2012zd} for exceptional underlying gauge groups. Furthermore, even assuming global chiral symmetry breaking, it remains to be seen if the breaking is to the maximal diagonal subgroup. We shall see that for certain phenomenologically relevant gauge theories, using first principle lattice simulations, there has been substantial progress to answer precisely these questions \cite{Catterall:2007yx,Catterall:2008qk,DelDebbio:2008wb,DelDebbio:2008zf,Catterall:2009sb,Hietanen:2009az,DelDebbio:2009fd,Kogut:2010cz,Karavirta:2011zg,Lewis:2011zb,Hietanen:2012qd,Hietanen:2012sz,Hietanen:2013fya,Hietanen:2013gva}.
We note that the classification of relevant underlying gauge theories for Technicolor models appeared in \cite{Dietrich:2006cm}, while a classification of the symmetry breaking patterns relevant for composite models of the Higgs as a pGB can be found in~\cite{Mrazek:2011iu,Bellazzini:2014yua}.

The first relevant observation is that the non-abelian global quantum unbroken flavor symmetries, for any underlying gauge theory with one Dirac species of fermions, are bound to be SU($2N_f$) or SU($N_f$)$\times$SU($N_f$), depending on whether the underlying fermion representation is (pseudo-)real or complex. 

An SU($2 N_f$) flavor symmetry can be achieved only if the new fermions belong to a real (like the adjoint) or pseudo-real (like the fundamental of SU(2) = Sp(2)) representation of the underlying gauge group.
In either case, both left-handed fermions and charge-conjugated right-handed anti-fermions can be recast, via a similarity transformation, to transform according to the same representation of the underlying fundamental gauge group. One can organize the two fermion components as a $2 N_f$ column with complex Weyl fermions indicated by the vector $\psi^f_c$ with $f = 1,..,2N_f$ and $c$ the new color index. 
 At this point, the simplest gauge invariant fermion bilinear we can construct  is $\psi^f_c \psi^{f'}_{c^{\prime}}$ with the color contracted either via a delta function for real representations or an antisymmetric epsilon term for pseudo-real ones. As fermions anticommute, a non vanishing condensate can be formed only if the full wave-function is completely antisymmetric with respect to spin, new color and flavor. Since Lorentz symmetry is conserved, the spin indices are contracted via an antisymmmetric tensor and therefore, according to the reality or pseudo reality of the representation we can have the following two patterns of chiral symmetry breaking:

\begin{itemize}
\item[Real:]  In this case we expect SU($2 N_f$)$\to$ SO($2 N_f$). The point is that the invariant product is symmetric (for instance $3 \times 3$ in SU(2)), and the flavor contraction must also be symmetric implying that the condensate belongs to the symmetric 2-index representation of SU($2 N_f$).
\item[Pseudo:] Here we expect SU($2 N_f$) $\to$ Sp($2 N_f$). In this case, as explained before, the invariant gauge singlet tensor is antisymmetric, therefore the condensate transforms as the antisymmetric 2-index representation of SU($2 N_f$).
\end{itemize}

Comparing these possible symmetry breaking patterns coming from a fundamental theory with fermions with the list of composite (pseudo-GB) Higgs possibilities, we can conclude that:
\begin{itemize}
\item[-] the minimal composite Goldstone Higgs scenario~\cite{MCH} cannot be realized in a simple minimal way~\cite{Serone}: in fact, it is based on SO(5)$\to$SO(4). However this chiral symmetry pattern cannot occur naturally because the minimal flavor symmetry SO(5) cannot be realized by an underlying fundamental fermionic matter theory.
\item[-] the next to minimal GB Higgs scenario is based on an enlarged SO(6) $\sim$ SU(4) global symmetry. The breaking SU(4)$\to$SO(4) is possible via a condensate belonging to the symmetric $10$ dimensional representation: however, such a breaking will generate 9 GBs belonging to a $(3,3)$ of SO(4), therefore no GB Higgs boson can be generated in the coset~\footnote{The SU(4)$\to$SO(4) breaking with two Higgs doublets used in~\cite{Mrazek:2011iu} is generated by an adjoint $15$ of SU(4), which however cannot be a condensate of Technifermions.}. On the other hand this pattern of chiral symmetry is extremely interesting for modern versions of minimal Technicolor models \cite{Sannino:2004qp,Dietrich:2006cm,Foadi:2007ue,Frandsen:2009mi} and their lattice studies \cite{Hietanen:2012sz,Hietanen:2013gva}. 
\item[-] the symmetry breaking SO(6)$\sim$SU(4)$\to$Sp(4)$\sim$SO(5) is also an interesting possibility both for (ultra) minimal Technicolor models \cite{Appelquist:1999dq,Duan:2000dy,Ryttov:2008xe} and the composite GB Higgs example \cite{Katz:2005au,Gripaios:2009pe,Galloway:2010bp,Barnard:2013zea,Ferretti:2013kya}, and for constructing UV completions of Little Higgs models \cite{Batra:2007iz}. Here the breaking is generated by an antisymmetric, with respect to the global flavor symmetry, $6$-dimensional representation, and the coset contains therefore 5 GBs. In terms of the SO(4) subgroup of SO(5), the GBs decompose into a $(2,2) + (1,1)$, thus also allowing, as we shall see, for a GB Higgs. In the following we will pursue this chiral symmetry breaking pattern which at the fundamental level is also being studied on the lattice \cite{Lewis:2011zb,Hietanen:2013fya}. 
\item[-] the next chiral symmetry breaking interesting pattern, from the composite GB Higgs boson point of view, is SU(5)$\to$SO(5). Here we have 14 GBs, decomposing as $(3,3) + (2,2) + (1,1)$ of SO(4)~\cite{Katz:2005au,Vecchi:2013bja,Ferretti:2013kya}.
\item[-] for SU(6) $\to$ Sp(6) we have two composite GB Higgs doublets and 6 singlets~\cite{Katz:2005au}.
\end{itemize}

From this list it is clear that, from the point of view of a fundamental theory with fermionic matter, the minimal scenario to investigate is SU(4)$\to$SO(5), for both a minimal Technicolor as well as composite GB Higgs scenario. The difference being in the way one embeds the electroweak theory within the global flavor symmetry. 
In the rest of the paper, we will discuss this possibility.

\section{A Minimal Fundamental Gauge Theory Setup: SU(2) with two Dirac Fermions} 
It is, in principle, possible to physically realize this pattern of chiral symmetry breaking starting from an underlying SU($2$)=Sp($2$) gauge theory with 2 Dirac flavors (i.e. four Weyl fermions) transforming according to the fundamental representation of the gauge group. We will discuss the lattice results \cite{Lewis:2011zb,Hietanen:2013fya} supporting the breaking of the global SU(4) symmetry to Sp(4) (locally isomorphic to SO(5)) via the formation of a non-perturbative fermion condensate in section \ref{link}. The underlying Lagrangian, in Dirac notation, is  
\begin{equation}\label{Lagrangian}
\mathcal{L} =  -\frac{1}{4}{F}_{\mu\nu}^a F^{a\mu\nu}
               + \overline{U}(i\gamma^{\mu}D_{\mu}-m)U
               + \overline{D}(i\gamma^{\mu}D_{\mu}-m)D \ ,
\end{equation}
where $U$ and $D$ are the two new fermion fields having a common bare mass
$m$, $F_{\mu\nu}^a$ is the field strength, and $D_\mu$ is the covariant
derivative.  The Dirac and SU(2) gauge indices of $U$ and $D$ are not
shown explicitly. Lattice simulations can extrapolate to the zero fermion mass case, and in that limit the Lagrangian has a global SU(4) symmetry corresponding to
the four chiral fermion fields
\begin{equation}
U_L=\frac{1}{2}(1-\gamma^5)U \ ,
~~~~~
U_R=\frac{1}{2}(1+\gamma^5)U \ ,
~~~~~
D_L=\frac{1}{2}(1-\gamma^5)D \ ,
~~~~~
D_R=\frac{1}{2}(1+\gamma^5)D \ .
\end{equation}
For $m\neq0$, the SU(4) symmetry is explicitly broken to a remaining Sp(4)
subgroup as follows.  The Lagrangian from (\ref{Lagrangian}) can be rewritten as
\begin{equation}\label{LagrangianQ}
\mathcal{L} =  -\frac{1}{4}{F}_{\mu\nu}^a F^{a\mu\nu}
               + i\overline{U}\gamma^{\mu}D_{\mu}U
               + i\overline{D}\gamma^{\mu}D_{\mu}D
               + \frac{m}{2}Q^T (-i\sigma^2) C \,EQ + { \frac{m}{2}\left(Q^T(-i\sigma^2)C\,EQ \right)^{\dagger} } \ ,
\end{equation}
where
\begin{equation}
Q = \left( \begin{array}{c} U_L \\ D_L \\ \widetilde{U}_L \\\widetilde{D}_L  \end{array} \right) \ ,
~~~~~~~~
E = \left(\begin{array}{cccc} 0 & 0 & 1 & 0 \\
      0 & 0 & 0 & 1 \\ -1 & 0 & 0 & 0 \\ 0 & -1 & 0 & 0 \end{array}\right) \ ,
\end{equation}
$C$ is the charge conjugation operator acting on Dirac indices, and
the Pauli structure $-i\sigma^2$ is the standard antisymmetric tensor acting on
color indices. We also have  $\widetilde{U}_L= -i\sigma^2C\overline{U}_R^T$ and  $\widetilde{D}_L = 
     -i\sigma^2C\overline{D}_R^T$. 
Under an infinitesimal SU(4) transformation defined by
\begin{equation}
Q \to \left(1+i\sum_{n=1}^{15}\alpha^nT^n\right)Q \ ,
\end{equation}
the Lagrangian (\ref{LagrangianQ}) becomes
\begin{equation}
\mathcal{L} \to \mathcal{L} + \frac{im}{2}\sum_{n=1}^{15}\alpha^nQ^T (-i\sigma^2) C \,
                              \left(ET^n+T^{nT}E\right)Q + {{\rm h.c.}}\ ,
\end{equation}
where $T^n$ denotes the 15 generators of SU(4) and $\alpha^n$ is a set of
15 constants.
The only generators that leave the Lagrangian invariant are those that obey
\begin{equation}
ET^n + T^{nT}E = 0
\end{equation}
which is precisely the definition of an Sp(4) Lie algebra.  From this, it is
straightforward to derive the ten Sp(4) generators in a specific
representation; see the appendix of \cite{Ryttov:2008xe}.

For $m=0$ the Lagrangian retains the full SU(4) symmetry but,
by analogy with the SU(3) theory of QCD, one might expect dynamical symmetry
breaking associated with the appearance of a nonzero vacuum expectation value,
\begin{eqnarray}
\langle \overline{U}U + \overline{D}D \rangle \neq0 \ .
\end{eqnarray}
Since this vacuum expectation value has the same structure as the terms
containing $m$ in the Lagrangian, the dynamical breaking would also be
SU(4)$\to$ Sp(4).
According to the Nambu-Goldstone theorem, the five broken generators are accompanied
by five GBs. 

This possible pattern of dynamical symmetry breaking must be checked
non-perturbatively via first-principles lattice simulations \cite{Lewis:2011zb,Hietanen:2013fya} and the results relevant for this work will be reported in section  \ref{link}. 

We also note that the choice of the antisymmetric vacuum  $E$ is not unique, meaning that we could have chosen another antisymmetric matrix also breaking SU(4) to Sp(4). However, the physical properties of the theory in isolation, i.e. before embedding the electroweak theory, such as the spectrum, decay constants and so forth, do not depend on this choice. 

Having set the notation for the two-color theory in isolation we turn our attention to the electroweak sector which we embed by assigning to the four Weyl fermions the following electroweak transformations:
\begin{itemize}
\item[-] Two first two Weyl fermions become one SU(2)$_L$ doublet $Q_L = (U_L,\, D_L)$ with zero hypercharge;
\item[-] Two SU(2)$_L$ singlets $\widetilde{U}_L$ and $\widetilde{D}_L$ with hypercharges $-1/2$ and $+1/2$ respectively.
\end{itemize}

This model has been considered in~\cite{Appelquist:1999dq,Duan:2000dy,Ryttov:2008xe} as a model for Technicolor and in~\cite{Galloway:2010bp} as a model with a pseudo-GB Higgs.~\footnote{A model based on SP(2N) has been studied in~\cite{Barnard:2013zea}, in the framework of the same global symmetry SU(4)/Sp(4).}

\section{Electroweak vacuum alignment for SU(4) $\to$ Sp(4)$\sim$ SO(5)}

We will start this analysis by considering a non-perturbative vacuum $\langle QQ  \rangle \propto \Sigma_0$ that, once the electroweak sector has been embedded, does not break the electroweak symmetry. As discussed in~\cite{Galloway:2010bp}, there are two electroweak inequivalent vacua, where by inequivalent one means that they cannot be related to one  another by an SU(2)$_L$ transformation, and they are:
\beq
\Sigma_A = \left( \begin{array}{cc}
i \sigma_2 & 0 \\
0 & i \sigma_2
\end{array} \right)\,, \qquad \Sigma_B = \left( \begin{array}{cc}
i \sigma_2 & 0 \\
0 & - i \sigma_2
\end{array} \right)\,, 
\eeq
where $\sigma_i$ are the Pauli matrices, and we wrote the matrix in a block form and chose the normalization to be real.
With either choice, the physical properties of the pGBs are the same:
in~\cite{Gripaios:2009pe}, the authors consider $\Sigma_A$ to build their model, while in~\cite{Galloway:2010bp} the model is constructed around $\Sigma_B$. 
In this paper, we will use $\Sigma_B$.

There is another alignment of the condensate which is of physical interest, given by the matrix
\beq
\Sigma_H = E  =\left( \begin{array}{cc}
0 & 1 \\
-1 & 0
\end{array} \right) \ .
\eeq
This vacuum completely breaks the electroweak symmetry, and can therefore be used to construct a Technicolor model~\cite{Appelquist:1999dq,Duan:2000dy,Ryttov:2008xe}.

\subsection{Vacua B and H}

Since the condensate transforms under SU(4) as $\Sigma \to u \Sigma u^T$ with $u \in$ SU(4), the unbroken generators of SU(4) are defined by
\beq
T^a \cdot \Sigma_B + \Sigma_B \cdot {T^a}^T = 0\,. 
\label{unbroken}
\eeq
The ten unbroken generators can be organized as follows:
\beq
S^{1,2,3} = \frac{1}{2} \left(  \begin{array}{cc}
\sigma_i & 0 \\
0 & 0
\end{array} \right)\,, \qquad S^{4,5,6} = \frac{1}{2} \left(  \begin{array}{cc}
0 & 0 \\
0 & - \sigma_i^T 
\end{array} \right)\,,
\eeq
which form an SU(2)$\times$ SU(2) subgroup of SO(5), while the remaining 4 are
\beq
S^{7,8,9} =  \frac{1}{2\sqrt{2}} \left(  \begin{array}{cc}
0 &i  \sigma_i \\
-i \sigma_i & 0
\end{array} \right)\,, \qquad S^{10} = \frac{1}{2\sqrt{2}} \left(  \begin{array}{cc}
0 &  1\\
1 & 0 
\end{array} \right)\,.
\eeq
The generators of the electroweak symmetry are identified with: $S^{1,2,3}$ for SU(2)$_L$ and $S^6$ for U(1)$_Y$.
The SU(2)$\times$SU(2) group generated by $S^{1, \dots 6}$ can therefore be thought of as the custodial symmetry of the SM Higgs potential.
The 5 broken generators, associated with the GBs are:
\beq
& X^1 = \frac{1}{2 \sqrt{2}} \left( \begin{array}{cc}
0 & \sigma_3 \\
\sigma_3 & 0
\end{array} \right)\,, \quad  X^2 = \frac{1}{2 \sqrt{2}} \left( \begin{array}{cc}
0 & i \\
-i & 0
\end{array} \right)\,, \quad  X^3 = \frac{1}{2 \sqrt{2}} \left( \begin{array}{cc}
0 & \sigma_1 \\
\sigma_1 & 0
\end{array} \right)\,,  & \\
&  X^4 = \frac{1}{2 \sqrt{2}} \left( \begin{array}{cc}
0 & \sigma_2 \\
\sigma_2 & 0
\end{array} \right)\,, \quad 
X^5 = \frac{1}{2 \sqrt{2}} \left( \begin{array}{cc}
1 & 0 \\
 0 & -1
\end{array} \right)\,. &
\eeq
Using the above decomposition, one can move in the quotient SU(4)/Sp(4) $ \sim $ SU(4)/SO(5) around the vacuum $\Sigma_B$ as follows:
\beq
\Sigma = e^{i \frac{\phi_i}{f} X^i} \cdot {\Sigma_B} \sim  {\Sigma_B} + \frac{1}{2 \sqrt{2} f} \left( \begin{array}{cccc}
0 & i \phi_5 & \phi_4 + i \phi_3 & \phi_2 - i \phi_1 \\
- i \phi_5 & 0 & - \phi_2 - i \phi_1 & \phi_4 - i \phi_3 \\
-\phi_4 - i \phi_3 & \phi_2 + i \phi_1 & 0 & i \phi_5 \\
- \phi_2 + i \phi_1 & -\phi_4 + i \phi_3 & -i \phi_5 & 0
\end{array} \right) + \mathcal{O} (\phi^2)\,. \nonumber \\
\eeq
The interactions of the GBs with the electroweak gauge bosons are obtained via the minimal coupling following the standard procedure. The associated Lagrangian term is:
\beq
\mbox{Tr} D_\mu \Sigma^\dagger D^\mu \Sigma \ . 
\eeq
However, given that the $S$ generators satisfy \eqref{unbroken}, this operator is unable to provide a mass to the observed massive SM gauge bosons. We have also chosen $S^6$ as the hypercharge generator in such a way that the electric charge generator is  $Q = T^3 + Y = S^3 + S^6$. It is straightforward to show that $Q$ satisfies the relation $Q \cdot \Sigma_H +  \Sigma_H \cdot Q^T=0$ with 
\beq
\Sigma_H = E  =\left( \begin{array}{cc}
0 & 1 \\
-1 & 0
\end{array} \right)\ , \qquad   \Sigma_H  = 2\sqrt{2}\, i\, X^4\cdot\Sigma_B \ .
\eeq
This implies that if $\phi_4$ acquires a non vanishing vacuum expectation value $\langle \phi_4 \rangle = v$, due to some yet unspecified dynamical source, then the electroweak symmetry breaks with $Q$ being the correct electric charge operator. Also, the fields $\phi^{1,2,3}$ are the GBs eaten by the massive $W$ and $Z$. The fluctuation around the vacuum of  $\phi_4$ is identified with the Higgs  $h$ and $\phi_5 = \eta$ is a singlet scalar. According to the analysis in \cite{Appelquist:1999dq} the Higgs-condensate, proportional to $\Sigma_H$, still leaves the following generators\footnote{The labeling of the generators is not identical, but very close, to the original terminology used in \cite{Appelquist:1999dq}.} unbroken:
\beq
S^1+{S^4}\,, \; S^2+{S^5}\,, \; S^3 + S^6\,, \; S^{7,9,10}\,, \; X^{1,2,3,5}\, ,
\label{unbroken-SM}
\eeq
while the broken ones can be written as
\beq 
S^1-{S^4}\,, \; S^2-{S^5}\,, \; S^3 - S^6\,, \; S^{8}\,, \; X^{4}\,.
\eeq
As already shown in \cite{Appelquist:1999dq}  the VEV along the direction $\Sigma_H$ breaks the SO(4) included in Sp(4) to an SU(2)$_D$, spanned by the first three generators in \eqref{unbroken-SM}, in agreement with the correct SM breaking pattern.  From the kinetic term of $\Sigma$ we can now determine the masses of the $W$ and $Z$ bosons.

\section{Phenomenology of the SU(4) $\to$ Sp(4)$\sim$SO(5) model}

Having analyzed the vacuum properties and the chiral symmetry breaking patterns relevant for the electroweak symmetry we move to investigate the various phenomenological aspects of the model. 
Several of the results in this section can be found in~\cite{Katz:2005au,Galloway:2010bp}: we will recall the main features, and stress on the connection between the Technicolor and pGB Higgs vacua.
We start by defining the vacuum of the theory as a superposition of the two vacua studied above
\beq
\Sigma_0 =\cos \theta\; \Sigma_B + \sin \theta\; \Sigma_H\, ,
\eeq
in such a way that $\Sigma_0^\dagger \Sigma_0 = 1$ is properly normalized.
The angle $\theta$ is, at this stage, a free parameter, which interpolates between a purely Technicolor model when $\theta=\pi/2$ to an unbroken phase for $\theta=0$, passing through a model of composite Higgs for small $\theta \ll 1$.
With the above condensate, the 5 broken generators are:
\beq
& Y^1  = c_\theta\, X^1 - s_\theta\, \frac{S^1-S^4}{\sqrt{2}}\,, \;  Y^2 = c_\theta\, X^2 + s_\theta\, \frac{S^2-S^5}{\sqrt{2}}\,, \;  Y^3 = c_\theta\, X^3 + s_\theta\, \frac{S^3-S^6}{\sqrt{2}}\,. & \\
& Y^4 = X^4\,, \; Y^5 = c_\theta\, X^5 - s_\theta\, S^{8}\,; &
\eeq
where $c_\theta = \cos \theta$ and $s_\theta = \sin \theta$.
The GBs that become the longitudinal components of  the $W$ and $Z$ gauge bosons are still  associated to the $Y^{1,2,3}$ generators. Working in the unitary gauge we use explicitly only the fields associated to $Y^{4,5}$ and write:
\beq
\Sigma = e^{\frac{i}{f} (h Y^4 + \eta Y^5)}\cdot \Sigma_0 = \left[ \cos \frac{x}{f}\; 1 + \frac{i}{x} \sin \frac{x}{f}\; \left( h Y^4 + \eta Y^5 \right) \right] \cdot \Sigma_0\ ,
\eeq
where $\displaystyle{x = \frac{\sqrt{h^2 + \eta^2}}{2\sqrt{2}}}$.
The kinetic term of $\Sigma$, upgraded to include the interactions with the gauge bosons via minimal coupling, yields:\beq
f^2 \,\, \mbox{Tr} (D_\mu \Sigma)^\dagger D^\mu \Sigma &=& \frac{1}{2} (\partial_\mu h)^2  + \frac{1}{2} (\partial_\mu \eta)^2 \nonumber \\ &&{ + \frac{1}{48 f^2} \left[ -  (h \partial_\mu \eta - \eta \partial_\mu h)^2\right]} + \mathcal{O} (f^{-3})   \nonumber \\
&  &+\left( 2 g^2 W^+_\mu W^{- \mu} + (g^2 + {g'}^2) Z_\mu Z^\mu \right) \left[ f^2 s_\theta^2 + \frac{s_{2 \theta} f}{2 \sqrt{2}} h \left(1-\frac{1}{12 f^2} (h^2 + \eta^2) \right)  \right. \nonumber \\
& &\left. + \frac{1}{8} (c_{2\theta} h^2 - s_\theta^2 \eta^2) \left( 1 -\frac{1}{{24} f^2} (h^2 + \eta^2) \right) + \mathcal{O} (f^{-3}) \right] \ .
\eeq
From the above expansion, we can identify the $W$ and $Z$ masses:
\beq
m_W^2 = 2 g^2 f^2 s_\theta^2\,, \qquad m_Z^2 = 2 (g^2 + {g'}^2) f^2 s_\theta^2 = m_W^2/c_W^2\,,
\eeq
thus $v = 2 \sqrt{2} f s_\theta$.
Furthermore, only the scalar $h$ couples singly to the massive gauge bosons, therefore it is the candidate to play the role of the Higgs boson. Its couplings are:
\beq
g_{h W W} &=& \sqrt{2} g^2 f s_\theta c_\theta = g m_W c_\theta = g_{hWW}^{SM} c_\theta\,, \\ 
g_{h Z Z} &=& \sqrt{2} (g^2 + {g'}^2) f s_\theta c_\theta =  \sqrt{g^2 + {g'}^2} m_Z c_\theta = g_{hZZ}^{SM} c_\theta\,, \\ 
g_{h h WW} &=& \frac{g^2 c_{2\theta}}{4}  = g_{hhWW}^{SM} c_{2 \theta}\,, \\
g_{h h ZZ} &=&  g_{h h WW}/c_W^2\,.
\eeq
The second scalar $\eta$ has couplings
\beq
g_{\eta \eta WW} &=& - \frac{1}{ 4} g^2 s^2_{\theta} = - g_{hhWW}^{SM} s_{\theta}^2\,, \\
g_{\eta \eta ZZ} &=&  g_{\eta \eta WW}/c_W^2\,.
\eeq
It is noteworthy that the kinetic term of $\Sigma$ is invariant under the parity transformation $\eta \to - \eta$, therefore $\eta$ is protected and will be stable. The non-topological GB Lagrangian respects a parity operation according to which the only possible terms must be even in the number of GBs.  This property has been used in~\cite{Frigerio:2012uc} to study $\eta$ as a composite dark matter candidate. However this apparent discrete symmetry is not a symmetry of the underlying theory. The breaking of this symmetry manifests itself at the effective Lagrangian level via topological-induced terms. These have been constructed explicitly for the chiral symmetry breaking pattern envisioned here in \cite{Duan:2000dy}. Here one finds also the correct gauging of the topological terms useful to consider the interaction with the gauge bosons.

\subsection{Loop induced Higgs potential}

While the dynamics does not have any preference to where the condensate is aligned in the SU(4) space, gauge interactions do because they only involve a subgroup of the flavor symmetry.
The same is true, as we will see, for the top Yukawa, or generically the mechanism that will generate a mass for the top.
The breaking of the flavor symmetry will then be communicated to the GBs via loops, which will therefore induce a potential determining the value of the angle $\theta$.
The loop-induced potential for this model has been computed in~\cite{Katz:2005au,Galloway:2010bp}, here we will simply recap the origin of the main components and discuss their physical interpretation.

\subsubsection{Gauge contributions}

The contribution to the one-loop potential of the gauge boson loops can be estimated by constructing the lowest order operator invariant under the flavor symmetry SU(4).
The gauge generators of SU(2)$_L$ are  $S^{1,2,3}$ while the one for U(1)$_Y$ is $S^6$.  Under a  vacuum rotation preserving the unbroken subgroup we have $S^i \to U S^i U^\dagger$ and the associated relevant  term for the effective potential coming from the gauge sector loop reads \cite{Peskin:1980gc,Preskill:1980mz}:
\beq
V_{SU2} &=& - C_g g^2 f^4 \sum_{i=1}^3 \mbox{Tr} \left(S^i \cdot \Sigma \cdot (S^i \cdot \Sigma)^\ast \right) \nonumber \\
 & \sim &  C_g g^2 \left( - \frac{3}{2} f^4 c_\theta^2 + \frac{3}{2 \sqrt{2}} f^3 c_\theta s_\theta h + \frac{3}{16} f^2 (c_{2\theta} h^2 - s_\theta^2 \eta^2) + \dots \right)\,;
\eeq
where $C_g$ is an unknown loop factor, and we have explicitly shown an expansion in powers of $f$ up to quadratic terms in the fields.
Analogously, for the contribution due to the U(1)$_Y$ generator we have:
\beq
V_{U1} &=& - C_g {g'}^2 f^4  \mbox{Tr} \left(S^6 \cdot \Sigma \cdot (S^6 \cdot \Sigma)^\ast \right) \nonumber \\
 & \sim &  C_g {g'}^2 \left( - \frac{1}{2} f^4 c_\theta^2 + \frac{1}{2 \sqrt{2}} f^3 c_\theta s_\theta h + \frac{1}{16} f^2 (c_{2\theta} h^2 - s_\theta^2 \eta^2) + \dots \right).
\eeq
To find the value of $\theta$ it is enough to minimize the field-independent term: $\partial V (\theta)/\partial \theta = 0$.
The constant $C_g$ encodes the loop factor, and it is expected to be positive: in this case, this part of the potential has a minimum for $\theta=0$, which therefore does not break the electroweak symmetry. Note also that the term with linear coupling of the ``Higgs'' $h$ is always proportional to the derivative of the potential, thus it is bound to vanish at the minimum.

\subsubsection{Top contribution}

To calculate the effects on the vacuum alignment induced by the top corrections we will follow the procedure established in~\cite{Peskin:1980gc,Preskill:1980mz}. Not having at our disposal a complete theory of flavor we assume that the top mass is generated via the following 4-fermion operator
\beq
\frac{y_t}{\Lambda_t^{{2}}} (Q t^c)^\dagger_\alpha \psi^T P^\alpha \psi
\eeq
where $\alpha$ is an SU(2)$_L$ index and the projectors $P^\alpha$ select the components of $\psi^T \psi$ that transform as a doublet of SU(2)$_L$ (i.e. the linearly transforming Higgs boson doublet properties). $\Lambda_t $ is some new dynamical scale.  
The projectors can be written as~\cite{Galloway:2010bp}
\beq
P^1 = \frac{1}{2} \left( \begin{array}{cccc}
0 & 0 & 1 & 0 \\
0 & 0 & 0 & 0 \\
-1 & 0 & 0 & 0 \\
0 & 0 & 0 & 0
\end{array} \right)\,, \qquad P^2 = \frac{1}{2} \left( \begin{array}{cccc}
0 & 0 & 0 & 0 \\
0 & 0 & 1 & 0 \\
0 & -1 & 0 & 0 \\
0 & 0 & 0 & 0
\end{array} \right)\,.
\eeq
When the techni-fermions condense this term generates the following operators:
\beq
y'_t  f (Q t^c)^\dagger_\alpha \mbox{Tr} (P^\alpha \Sigma) \sim - y'_t  \left( f s_\theta + \frac{1}{2 \sqrt{2}} c_\theta h  - \frac{1}{16 f} s_\theta (h^2 + \eta^2) + \dots \right) t_R t_L^c
\eeq
Here $y'_t$ is proportional to $y_t (4\pi f)^2/\Lambda^2_t$. We have not assumed the underlying fermionic dynamics to be near conformal. If this were the case the relation changes as it is the case for walking Technicolor. 
The first term in the expansion generates a top mass $m_{top} = y'_t f s_\theta$ when $\theta \neq 0$, and the coupling of the Higgs to the top is
\beq
y_{h\bar{t}t} = \frac{y'_t c_\theta}{2 \sqrt{2}} = \frac{m_{top}}{v} c_\theta\,.
\eeq
From the form of the operator above, the contribution of the top loop can be estimated as
\beq
V_{top} &=& - C_t {y'_t}^2 f^4 \sum_{\alpha=1}^2 \left[  \mbox{Tr} (P^\alpha \Sigma) \right]^2 \nonumber \\
 & \sim & - C_t  {y'_t}^2 \left[f^4 s_\theta^2 + \frac{1}{\sqrt{2}} f^3 c_\theta s_\theta h + \frac{1}{8} f^2 (c_{2\theta} h^2 - s_\theta^2 \eta^2) + \dots \right]
 \eeq
where again we expect the coefficient $C_t$ to be positive. 
In this case, the minimum is located at $\theta = \pi/2$, which would break the electroweak symmetry at the condensate scale. The vacuum preferred by the top corrections therefore corresponds to the standard Technicolor-like limit \cite{Dietrich:2006cm,Appelquist:1999dq,Ryttov:2008xe}. 

This Technicolor vacuum limit is quite interesting: in fact, one would have that all the linear couplings of $h$ to gauge bosons and to the top vanish, the reason being that in this limit an extra $U(1)$ symmetry remains unbroken upon gauging the electroweak symmetry. This global U(1) symmetry is reminiscent of the QCD-like underlying techni-baryon symmetry linking $h$ and $\eta$ into a complex di-techni-quark GB. Intriguingly a similar decoupling property for the would be composite Higgs GB was observed in the Hosotani model \cite{Hosotani:2009jk}. Here too, in the decoupling limit, the near decoupled state becomes a dark matter candidate. 
This property of the Technicolor vacuum has been used extensively for dark matter model building \cite{Gudnason:2006ug,Gudnason:2006yj,Nardi:2008ix,Ryttov:2008xe,Foadi:2008qv,Frandsen:2009mi,DelNobile:2011je} and it is supported also by recent pioneering lattice investigations \cite{Lewis:2011zb,Hietanen:2012sz,Hietanen:2013fya}. In the Technicolor limit, the pGB $h$ ceases to be a Higgs-like particle: the physical Higgs-like state now becomes the lightest techni-flavor (and SM) singlet composite scalar state associated with the fluctuations of the condensate orthogonal to the GB directions. The coupling of this state to the gauge bosons and fermions does not vanish for $\theta=\pi/2$: we will investigate the properties of this state in the following section.

In this vacuum, as explained before, $\eta$ and $h$ are degenerate and acquire the following loop-induced mass term:
\beq
{ m^2_{DM}} = m_h^2 = m_\eta^2 = \frac{f^2}{4} \left( C_t {y'_t}^2 - C_g \frac{3 g^2 + {g'}^2}{2}  \right)\,.
\eeq
 There is, in principle, another possible contribution to the mass of the dark matter candidate above coming from a different source of explicit SU(4) symmetry breaking discussed in \cite{Ryttov:2008xe}. This breaking term reinforces the alignment of the vacuum in the Technicolor direction. Interestingly the weak interactions like to misalign the Technicolor vacuum while the top corrections tend to re-align the vacuum in the Technicolor direction providing a positive mass term to the natural dark matter candidate $h + i \eta$. This state is therefore a complex pGB. The top corrections to the di-quark GB state were not included in \cite{Dietrich:2009ix}. 

\subsubsection{Explicit breaking of SU(4)}
\label{ESBSU}

Another source for the Higgs potential are eventual terms that break explicitly SU(4). Sources that do not upset the $\theta=\pi/2$ vacuum were constructed in  \cite{Gudnason:2006yj,Foadi:2007ue,Ryttov:2008xe} assuming natural breaking of the SU(4) symmetry via four-fermion interactions. However in~\cite{Katz:2005au}, for minimal models of composite (Goldstone) Higgs,  such a term is added ad-hoc to give mass to $\eta$; in~\cite{Galloway:2010bp} they are generated by gauge invariant masses of the techni-fermions.
In both cases, the results are the same: here we will follow the idea that such a term is generated by the explicit SU(4) violating masses of the techni-fermions.
As we want such masses to be invariant under the gauged symmetry, we can assume that the mass term is aligned with the condensate $\Sigma_B$, so that $M = \mu \Sigma_B$.
In this case, the contribution to the potential can be written as~\cite{Galloway:2010bp}:
\beq
V_{m} &=& C_m f^4 \mbox{Tr} (\Sigma_B \cdot \Sigma)\nonumber \\
& \sim & C_m \left( - 4 f^4 c_\theta + \sqrt{2} f^3 s_\theta h + \frac{1}{4} f^2 c_\theta (h^2 + \eta^2) + \dots \right)
\eeq
Note that contrary to the gauge and top loops, the coefficient $C_m$ is not expected to be positive and can have both signs.
This potential term contributes to off-set the ground state from $\theta = \pi/2$.
Defining
\beq
X_t = {y'_t}^2 C_t - \frac{3 g^2 + {g'}^2}{2} C_g\,, \qquad \mbox{and} \quad X_m = C_m\,,
\eeq
the potential to minimize reads, up to an overall $f^4$,
\beq
V (\theta) = X_t c_\theta^2 - 4 X_m c_\theta + \mbox{constants}
\eeq
which is extremized for
\beq
\theta = 0\,, \quad c_\theta = \frac{2 X_m}{X_t}\,.
\eeq
The latter solution, which corresponds to broken EWS, is possible only for $X_t > 2 |X_m|$, in which case it is indeed the minimum of the potential.
In the other case, the only allowed solution is $\theta=0$.
The masses for the scalar fields are given by
\beq
m_h^2 &=& \frac{f^2}{4} (X_t (1-2 c_\theta^2) + 2 X_m c_\theta)\,, \\
m_\eta^2 &=& \frac{f^2}{4} (X_t (1- c_\theta^2) + 2 X_m c_\theta)\,.
\label{mass-ESB}
\eeq

On the solution $c_\theta = 2 X_m/X_t$, the masses read
\beq
m_h^2 &=& \frac{f^2}{4} \frac{X_t^2 - 4 X_m^2}{X_t} = \frac{f^2}{4} s_\theta^2 X_t\,, \\
m_\eta^2 &=& \frac{f^2}{4} X_t\,.
\label{mass-ESB}
\eeq
We see that $m_h^2 > 0$ for $X_t > 2 |X_m|$, for which there is always a solution for $c_\theta$. Furthermore, we recover the relation $m_\eta^2 = m_h^2/s_\theta^2$~\cite{Galloway:2010bp}.

On the other solution $\theta=0$, where the electroweak symmetry is unbroken, the masses read
\beq
m_h^2 = \frac{f^2}{4} (2 X_m - X_t)\,, \qquad m_\eta^2 = \frac{f^2}{2} X_m\,.
\eeq
This solution is a stable minimum for $X_t < 2 X_m$.

\subsubsection*{More on the Higgs mass and fine tuning}

Using the expressions for the top and gauge masses, the composite GB Higgs can be rewritten as
\beq
m_h^2 = \frac{f^2 s_\theta^2}{4} \left( {y'_t}^2 C_t - \frac{3 g^2 + {g'}^2}{2} C_g \right) = \frac{C_t m_t^2}{4} \left( 1 - \frac{2 m_W^2 + m_Z^2}{4 m_t^2} \frac{C_g}{C_t} \right) \sim  \frac{C_t m_t^2}{4} \left( 1 - 0.18 \frac{C_g}{C_t} \right)\,. \nonumber \\
\eeq
This shows explicitly that the contribution of the gauge loops is typically smaller than the top one, even assuming $C_g \sim C_t$.
Also, neglecting the gauge contribution, to fit the observed Higgs mass $m_h = 125$ GeV, we would need
\beq
C_t \sim 2\,.
\eeq
Another interesting point is related to the value of $\theta$: in fact, this angle parametrizes the corrections to the Higgs couplings to the gauge bosons, which are by now well constrained by LHC data. Therefore, a realistic model would require $\theta$ to be small. In order to achieve this, we would need:
\beq
c_\theta \sim 1 \quad \Rightarrow \quad X_t \sim 2 X_m\,.
\eeq
In other words, a realistic model would require a non-trivial fine tuning between the contribution of the top loops, and the contribution of the explicit breaking of SU(4), which is induced by a completely different mechanism.

\subsubsection{Variation on the theme}

We can now explore some variations of the above scenario. For example we can gauge an extra U(1). The only possibility is to gauge the symmetry generated by $X^5$ which commutes with both SU(2)$_L$ and U(1)$_Y$.
The new gauge boson $X^\mu$ will contribute to the effective potential as follows
\beq
V_{X} &=& - C_g g_X^2 f^4  \mbox{Tr} \left(X^5 \cdot \Sigma \cdot (X^5 \cdot \Sigma)^\ast \right) \nonumber \\
 & \sim &  C_g g_X^2 \left( \frac{1}{2} f^4 (2 c_\theta^2-1) - \frac{1}{\sqrt{2}} f^3 c_\theta s_\theta h - \frac{1}{8} f^2 (c_{2\theta} h^2 - s_\theta^2 \eta^2) + \dots \right)
\eeq
Although this contribution has the familiar form of the contribution coming from the electroweak gauge terms, the preferred minimum is at $c_\theta = 0$.  This happens because the Technicolor ground state does not break the U(1)$_X$ symmetry thus leaving $X_\mu$ massless. In fact:
\beq
m_X^2 = 4 g_X^2 f^2 c_\theta^2\,.
\eeq
The net effect of this contribution would therefore be to add to the top loop.

Another variant is to generate a mass for the top via a heavy mediator $\Psi$.
The idea is to complement the theory with a new fermion belonging to a complete representation of SU(4), and couple it to $\Sigma$ in an SU(4) invariant way; the mass is then communicated to the top sector by an SU(4) violating mixing term of the form~\cite{Katz:2005au}
\beq
\lambda_1 f \bar{\Psi} \Sigma \Psi + \lambda_2 f Q_\psi Q + \lambda_3 f T_\psi t^c
\eeq
where $Q_\psi$ and $T_\psi$ are components of $\Psi$ with the same quantum numbers as the quark doublet and the right-handed top singlet.
However, it is not possible to find a representation of SU(4) which contains a doublet and a singlet with the correct hypercharge, following the embedding of the hypercharge generator discussed above.
One may think of embedding the U(1)$_Y$ as a superposition of the two possible U(1)'s, i.e. $Y = c_\alpha S^6 + s_\alpha X^5$.
However, there is no vacuum that can preserve both $S^6$ and $X^5$ and as a consequence there would be no QED unbroken U(1).
This is interesting as it allows us to discard the construction used in~\cite{Katz:2005au} to generate the top mass.
In \cite{Gripaios:2009pe}, the global flavor symmetry is extended to SU(4)$\times$U(1) and the hypercharge is identified with a linear combination of $S^6$ and the external U(1): this is a trick commonly used in models of composite pseudo-GB Higgs.
However, it is very unlikely that the condensate in any realistic fundamental theory, featuring fundamental vector-like fermionic matter, can break SU(4)$\times$U(1)$\to$SO(5).
In fact, one could imagine to generate the extra U(1) by adding a single fermion in the adjoint representation, which carries hypercharge.
However, the two sectors of fermions do not talk to each other, and the dynamics will generate, at first, two condensates: one breaking SU(4) and the other breaking U(1) independently. Gauge dynamics for vector-like theories with several fermionic matter fields were investigated in~\cite{Ryttov:2009yw,Chen:2010er}. To achieve the desired symmetry breaking pattern, for these theories, one would need to introduce fields transforming simultaneously under the two flavor symmetries. Fundamental scalar fields can easily accommodate this patterns at the price of introducing unnatural scenarios. On the other hand, chiral gauge theories (where a mass term for the fermions is prohibited) can break simultaneously different global symmetries as well as the underlying gauge dynamics and would constitute and interesting avenue to explore~\cite{Appelquist:2000qg}.

\section{Introducing the Techni-Higgs (the $\sigma$)}

As already mentioned above, the dynamics contains a would-be Higgs boson, besides the GB of the flavor symmetry breaking, which behaves like the $\sigma$ particle in QCD and is a singlet under the flavor symmetry.
To study its effect, we add its contribution to the potential for the GBs:
\beq
\mathcal{L} = \frac{1}{2} (\partial_\mu \sigma)^2 - \frac{1}{2} \kappa_M (\sigma) M^2 \sigma^2 + \kappa_G (\sigma)\; f^2 \,\, \mbox{Tr} (D_\mu \Sigma)^\dagger D^\mu \Sigma + \kappa_t (\sigma)\; y'_t  f (Q t^c)^\dagger_\alpha \mbox{Tr} (P^\alpha \Sigma)\ , \label{eq:Lsigma}
\eeq
where we have introduced the field $\sigma$ with a potential term, together with the modified GB kinetic term and the top Yukawa operator.
The dynamics generates the couplings of $\sigma$ to the above operators, encoded into the functions $\kappa_{M,G,t}$.
The Lagrangian is well defined when the space-time fluctuations of the field $\sigma$ are small compared to the scale of the dynamics $4\pi f$. De facto $\sigma$ becomes a background field. In this case one can indeed safely neglect terms with higher derivatives acting on $\sigma$, and naturally assume that the functions $\kappa$ can be expanded for small values of $\sigma$ as
\beq
\kappa_{M,G,t} (\sigma) \sim 1 + k^{(1)}_{M,G,t} \frac{\sigma}{4 \pi f} + \frac{1}{2} k^{(2)}_{M,G,t} \frac{\sigma^2}{(4 \pi f)^2} + \dots \ ,
\eeq
with the coefficients $k^{(n)}$ of order unity. We can therefore use the Lagrangian in \eqref{eq:Lsigma} to determine the loop-induced potential for the GBs:
\beq
V_{\rm 1-loop} = \kappa_G(\sigma)\; (V_{SU2} + V_{U1}) + \kappa_t (\sigma)^2\; V_{top} + \kappa_m (\sigma)\; V_m\ ,
\eeq
where $\kappa_m$ is a new function associated with the SU(4) explicit breaking term.
The potential above, augmented by the intrinsic $\sigma$ potential term (with its squared mass term factored out), needs to be minimized with respect to the angle $\theta$ parametrizing a given vacuum choice, as well as $\sigma$. It is convenient to split $\sigma$ into a vacuum expectation value and the slowly fluctuating field $\varphi$ as follows $\sigma = \sigma_0 + \varphi$.  $\varphi$ is identified with the massive physical degree of freedom, i.e. the techni-Higgs.
For the angle theta, the minimization condition is similar to the one performed in the previous sections, up to appropriate factors of $\kappa_X (\sigma_0)$:
\beq
\cos \theta = \frac{2 \kappa_m(\sigma_0)\, C_m}{{y'_t}^2 \kappa_t^2 (\sigma_0)\, C_t - \frac{3 g^2 + {g'}^2}{2} \kappa_G (\sigma_0)\, C_g}\,.
\eeq
To determine $\sigma_0$, on the other hand, we need to solve:
\beq
\frac{M^2}{f^4} \left( \kappa_M  \sigma_0 + \frac{1}{2} \kappa'_M  \sigma_0^2 \right) - 2 C_t {y'_t}^2 s^2_\theta \kappa_t \kappa'_t - C_g \frac{3 g^2 + {g'}^2}{2} c^2_\theta  \kappa'_G - 4 C_m c_\theta \kappa'_m = 0\,,
\eeq
where all the $\kappa$ functions and their derivatives with respect to $\sigma$ are evaluated at $\sigma_0$.
Consistency requires $\sigma_0\ll 4 \pi f$ for the Taylor expansion of the $\kappa$ functions to be valid.

Furthermore, from \eqref{eq:Lsigma} we see that the $\kappa$ functions also encode the couplings of the techni-Higgs to gauge bosons and the top. In fact, expanding around $\sigma_0$ we find:
\beq
& m_W^2 = \kappa_G (\sigma_0) \cdot 2 g^2 f^2 s^2_\theta\,, \qquad \mbox{and} \quad g_{WW\varphi} = \frac{\kappa'_G (\sigma_0)}{\kappa_G (\sigma_0)} m_W^2\ , & \\
& m_t = \kappa_t (\sigma_0) \cdot y'_t f s_\theta\,, \qquad \mbox{and} \quad g_{t\bar{t} \varphi} = \frac{\kappa'_t (\sigma_0)}{\kappa_t (\sigma_0)} m_t\,. &
\eeq
These relations become relevant when identifying the techni-Higgs as the SM Higgs.

\subsection{The Technicolor vacuum}

Let us now focus on the Technicolor vacuum, i.e. $\theta = \pi/2$, which is obtained in the limit $C_m = 0$.
In this case the SM Higgs can only be played by the techni-Higgs scalar $\varphi$, while the two pseudo-GBs $h$ and $\eta$ become degenerate. The associated complex state is stable and can play the role a complex dark matter state. Phenomenology requires the couplings of $\varphi$ to the gauge bosons to be close to the SM ones, yielding the following constraints on the function $\kappa_G$:
\beq
\frac{\kappa'_{G} (\sigma_0)}{\kappa_G (\sigma_0)} \sim \frac{k^{(1)}_G}{4 \pi f} \approx \frac{2}{v} = \frac{1}{\sqrt{2} f} \Rightarrow k^{(1)}_G \sim 2  \sqrt{2} \pi\,,
\eeq
where we are neglecting higher order terms in $\sigma_0/(4 \pi f)$.
The same analysis for the top would imply
\beq
\frac{\kappa'_t (\sigma_0)}{\kappa_t (\sigma_0)} \sim \frac{k^{(1)}_t}{4 \pi f} \approx \frac{1}{v} = \frac{1}{2\sqrt{2} f} \Rightarrow k^{(1)}_t \sim \sqrt{2} \pi\,. \label{eq:tH2top}
\eeq

We now consider the various contributions to the physical mass for the would-be Higgs $\varphi$, and  to the one of the complex dark matter particle formed by the two GBs.
Truncating the Taylor expansion of the $\kappa$ function to the first relevant orders we find:
\beq
m_\varphi^2 &=& M^2 \left( 1 + 3 k_M^{(1)} \frac{\sigma_0}{4 \pi f} \right) - \frac{\left((k_t^{(1)})^2 +k_t^{(2)} \right)}{2 \pi^2} \frac{C_t {y'_t}^2 f^2}{4}\,, \\
m_{DM}^2 &=&  \frac{C_t {y'_t}^2 f^2}{4} \left[ 1 - \frac{3 g^2 + {g'}^2}{2 {y'_t}^2} \frac{C_g}{C_t} + \left( 2 k_t^{(1)} - \frac{k_G^{(1)} C_g}{C_t} \frac{3 g^2 + {g'}^2}{2 {y'_t}^2} \right) \frac{\sigma0}{4 \pi f} \right]  
\,.
\eeq
In our calculation, $\sigma_0 \ll 4 \pi f$, therefore, at leading order the dark matter mass is the same as we obtained in the previous section
\beq
m_{DM}^2 \sim \frac{C_t {y'_t}^2 f^2}{4}  \simeq \frac{C_t}{4}m_t^2\,,
\eeq
where $m_t = y'_t  f$ for $\theta = \pi/2$ and $\kappa_t (\sigma_0) \sim 1$. 
Analogously, neglecting terms suppressed by $\sigma_0/(4 \pi f)$, the techni-Higgs mass is given by
\beq
m_{\varphi}^2 \simeq M^2 - \frac{\left((k_t^{(1)})^2 +k_t^{(2)} \right)}{2 \pi^2}  m_{DM}^2 \sim M^2 - C_t\frac{\left((k_t^{(1)})^2 +k_t^{(2)} \right)}{8 \pi^2} m_t^2 \ .
\eeq
The above formula provides a nice correlation between the would-be Higgs mass, the mass of the dark matter candidate and the top mass under the assumption that no other explicit SU(4) breaking terms are present in the theory. 
In particular, this formula shows the possibility that a potentially large value of $M$ generated by the strong dynamics, may be cancelled by the mass of the dark matter candidate, which is generated by the top loop as suggested in \cite{Foadi:2012bb}.

Finally, we need to check the consistency of our calculation by checking the value of $\sigma_0$, and making sure that it is not too big to invalidate the expansion. An approximate solution for $\sigma_0$ reads:
\beq
\sigma_0 \sim \frac{4 \pi f^3 k_t^{(1)} C_t {y'_t}^2}{8 \pi^2 M^2 - \left((k_t^{(1)})^2+ k_t^{(2)} \right) C_t {y'_t}^2 f^2} \sim 2 \sqrt{2} f \cdot \frac{k_t^{(1)}}{\sqrt{2} \pi} \frac{m_{DM}^2}{m_\varphi^2}\,.
\eeq
From this equation we see that a small correction $\sigma_0 \ll 4 \pi f$ would require either $m_{DM} < m_\varphi$, or small $k_t^{(1)} < \sqrt{2} \pi$ thus implying that the techni-Higgs has a coupling to the top smaller that the SM expectation (see Eq.(\ref{eq:tH2top})). Nevertheless, this analysis does not exclude the possibility to achieve a light techni-Higgs within a consistent dynamical framework. Furthermore the resulting value of the techni-Higgs mass can be further lowered because the intrinsic value of $M$ itself can be small with respect to $4\pi f$. The most discussed example in the literature is the one according to which $M$ is reduced because the underlying dynamics is near conformal. In this case the techni-Higgs is identified with the techni-dilaton~\cite{Yamawaki:1985zg,Dietrich:2005jn,Appelquist:2010gy}. Several model computations have been used to estimate $M$ ranging from the use of the truncated Schwinger-Dyson equations~\cite{Gusynin:1989jj,Holdom:1986ub,Holdom:1987yu,Harada:2003dc,Kurachi:2006ej,Doff:2008xx,Doff:2009nk,Doff:2009kq,Doff:2009na}  to computations making use of orientifold field theories~\cite{Sannino:2003xe,Hong:2004td}. Perturbative examples have proven useful to demonstrate the occurrence of a calculable dilaton   state parametrically lighter than the other states in the theory \cite{Grinstein:2011dq,Antipin:2011aa,Antipin:2012sm}. Last but not the least recent first principle lattice simulations \cite{Fodor:2014pqa,Fodor:2012ni} support this possibility for certain fundamental gauge theories  put forward in \cite{Hong:2004td,Dietrich:2005jn,Dietrich:2006cm}.

\subsection{The pseudo-Goldstone Higgs vacuum and beyond}

We now consider the general $\theta$ case, including also the explicit SU(4) breaking term, and determine the masses of the 3 scalars.  The $\eta$ state does not mix, to the quadratic order, with the other two states, while $h$ and $\varphi$ do mix with each others. To the lowest order in $\sigma_0/(4\pi f)$ we obtain:
\beq
m_\eta^2 & = & \frac{f^2}{4} \left[ C_t {y'_t}^2 - C_g \frac{3 g^2 + {g'}^2}{2} \right]
\,, \\
m_h^2 &=& m_\eta^2 s_\theta^2\,, \\
m_\varphi^2 &=& M^2  - \frac{\left((k_t^{(1)})^2 +k_t^{(2)} \right) s_\theta^2 + k_m^{(2)} c_\theta^2}{2 \pi^2} C_t {y'_t}^2 \frac{f^2}{4} - \frac{k_G^{(2)}   - 2  k_m^{(2)}}{4 \pi^2} c_\theta^2 C_g \frac{3 g^2 +{g'}^2}{2} \frac{f^2}{4} \,, \\
m_{h \varphi}^2 &=& - \frac{f^2\,s_{2\theta}}{8\pi \sqrt{2}}  \left[ (2 k_t^{(1)} - k_m^{(1)}) C_t {y'_t}^2 - (k_G^{(1)} - k_m^{(1)} ) C_g \frac{3 g^2 + {g'}^2}{2} \right]\,. 
\eeq
It is possible to further simplify the above equations by dropping the contribution of the gauge loops which are typically smaller than the top one. We get
\beq
m_\eta^2 &=& \frac{f^2}{4} C_t {y'_t}^2\,, \\
m_h^2 &=& m_\eta^2 s_\theta^2\,, \\
m_\varphi^2 &=& M^2 - m_\eta^2 \left( (\xi_t^2 + \xi_t^{(2)})  s_\theta^2 + \xi_m^{(2)} c_\theta^2 \right)\,, \\
m_{h \varphi}^2 &=& - (\xi_t - \xi_m) m_\eta^2 s_{2\theta}\,;
\eeq
where we have defined
\beq
\xi_t = \frac{k_t^{(1)}}{\sqrt{2} \pi} = \frac{g_{t\bar{t}\varphi}}{g^{SM}_{t\bar{t}h}}\,, \qquad \xi_t^{(2)} = \frac{k_t^{(2)}}{2 \pi^2}\, ,  \qquad \xi_m = \frac{k_m^{(1)}}{2 \sqrt{2} \pi}\,, \qquad \xi_m^{(2)} = \frac{k_m^{(2)}}{2 \pi^2}\,.
\eeq
The mass eigenvalues are given by:
\begin{multline}
m_{h_{1,2}}^2 = \frac{1}{2} \left[ M^2 - m_\eta^2 \left( \xi_m^{(2)} c_\theta^2 + (\xi_t^2 + \xi_t^{(2)}-1) s_\theta^2\right) \mp \right.\\
\left. \sqrt{[M^2 - m_\eta^2 (\xi_m^{(2)} c_\theta^2 + (\xi_t^2 + \xi_t^{(2)} + 1) s_\theta^2)]^2 + 4 m_\eta^4 (\xi_t - \xi_m)^2 s_{2\theta}^2} \right]\,.
\end{multline}
It is instructive to investigate the limit of small $\theta$, typically needed in models of composite pseudo-GB Higgs where $M \sim m_\eta \sim f \gg v \sim 2 \sqrt{2} f\, \theta$. We deduce
\beq
m_{h_2}^2 &\sim& M^2 - m_\eta^2 \xi_m^{(2)}\,, \\
m_{h_1}^2 &\sim& m_\eta^2 \theta^2 \left( 1- 4 \frac{m_\eta^2}{m_{h_2}^2} (\xi_t- \xi_m)^2 \right)\,.
\eeq
The mixing angle $\alpha$ between the two states, in the same approximation of small $\theta$, is
\beq
\tan \alpha \sim \alpha \sim 2 \frac{m_\eta^2}{m_{h_2}^2} (\xi_t - \xi_m) \theta\,.
\eeq
This angle can have a relevant impact on the phenomenology of the Higgs boson. Let's assume, in fact, to identify the discovered Higgs with the lightest state $h_1$. Its couplings must be close to the SM ones. However, the couplings of $h_1$ will get a contribution from the couplings of $\varphi$ via the mixing angle $\alpha$. After having defined the quantity $\xi_G = \frac{k_G^{(1)}}{2 \sqrt{2} \pi}$, the couplings of $\varphi$ to the $W$'s and top are given by $\xi_G s_\theta$ and $\xi_t s_\theta$ times the SM values.
We have
\beq
& \frac{g_{WWh_1}}{g^{SM}_{WWh}} = c_\theta c_\alpha + \xi_G s_\theta s_\alpha \sim 1 - \left(\frac{1}{2} - \xi_G \, 2 \frac{m_\eta^2}{m_{h_2}^2} (\xi_t - \xi_m) + 2 \frac{m_\eta^4}{m_{h_2}^4} (\xi_t-\xi_m)^2 \right) \theta^2 + \mathcal{O} (\theta^4)\,, & \\
& \frac{g_{t\bar{t}h_1}}{g^{SM}_{t\bar{t}h}} = c_\theta c_\alpha + \xi_t s_\theta s_\alpha \sim 1 - \left( \frac{1}{2} - \xi_t \, 2 \frac{m_\eta^2}{m_{h_2}^2} (\xi_t - \xi_m) + 2 \frac{m_\eta^4}{m_{h_2}^4} (\xi_t-\xi_m)^2 \right) \theta^2+ \mathcal{O} (\theta^4)\, .&
\eeq
The mixing with $\varphi$ has generated a correction of order $\theta^2$ to the couplings, at the same order as the contribution of the pseudo-GB nature of the Higgs arises at the $\theta^2$ level.  Hence the bounds on $\theta$, and therefore the required fine tuning between the top loop and the explicit SU(4) breaking, are sensitive to the presence of the $\sigma$ state and its correction cannot, in general, be neglected. 

As a consistency check one can determine again  $\sigma_0$ (neglecting also the gauge loops) and one finds
\beq
\sigma_0 &\sim& \frac{f^3 C_t {y'_t}^2}{2 \pi}  \frac{k_t^{(1)} s_\theta^2 + k_m^{(1)} c_\theta^2}{M^2 + \frac{f^2 C_t {y'_y}^2}{8 \pi^2} \left( ((k_m^{(1)})^2- 2 k_m^{(1)} k_t^{(1)} - k_m^{(2)})c_\theta^2 - ((k_t^{(1)})^2 + k_t^{(2)}) s_\theta^2 \right)} \nonumber \\
& \sim & 2 \sqrt{2} f  \frac{2 \xi_m m_\eta^2}{m_{h_2}^2 + 4 \xi_m (\xi_m - \xi_t) m_\eta^2} \,.
\eeq
therefore, for consistency, we shall either have small $\xi_m$, or $m_\eta \ll M$.

 \section{Chiral Symmetry breaking and predictions for the Spin One Spectrum from the Lattice}
\label{link}

Without the specific knowledge of an underlaying gauge theory it is impossible to provide a reasonable prediction for the energy scale when the spin-one spectrum of the theory will become accessible at colliders. The coefficients of the effective Lagrangians are all unknown allowing, at best, order unity predictions for low energy processes. 
Furthermore, for the phenomenological analyses presented above, whether they were meant for composite Higgs models of pseudo-GB or non-GB nature, it was assumed that a specific pattern of chiral symmetry breaking, namely SU(4) breaking to Sp(4)$\sim$ SO(5), occurs via a more fundamental dynamics. This pattern was assumed to occur before embedding the electroweak sector, before extending the model in order to provide masses to the SM fermions and, last but not the least, before the further explicit breaking of the original SU(4) symmetry. The conspiracy of these different ingredients is required to provide a successful model of electroweak symmetry breaking and fermion mass generation.

The first question to answer is therefore: 
Does it exist a fundamental gauge theory example, not suffering from quadratic divergences, that supports the dynamical breaking of SU(4) to Sp(4)$\sim$ SO(5)?  

The answer, from lattice simulations, is positive \cite{Lewis:2011zb,Hietanen:2013fya}. The dynamics studied is indeed an SU(2) gauge theory with two dynamical Dirac fermions transforming according to the fundamental representation of the gauge group. Using the Wilson formulation of the lattice action in \cite{Lewis:2011zb} clear signs, further analyzed in \cite{Hietanen:2013fya}, of chiral symmetry breaking have appeared. They consist in having shown that the GB squared mass vanishes linearly with the underlying common fermion masses, that the GB decay constant $f$ divided by $Z_a$ (the axial current renormalisation factor\footnote{The perturbative lattice determination of $Z_a$ and the vector renormalisation factor $Z_v$ for any SU(N) gauge theory and any matter representation can be found in \cite{DelDebbio:2008wb}.}), does not vanish in the chiral limit. Furthermore the spectrum of spin-one vector and axial resonances is well separated from the GBs with the ratio of the spin-one masses to the GB masses growing to infinity towards the chiral limit. 

We now provide the local operators, and the associated two-point correlators, studied on the lattice \cite{Lewis:2011zb,Hietanen:2013fya} in the Dirac notation. For the QCD-like meson degrees of freedom the local operators are
\begin{eqnarray}
{\cal O}_{\overline{U}D}^{(\Gamma)}(x) &=& \overline{U}(x)\Gamma D(x) \,, \\
{\cal O}_{\overline{D}U}^{(\Gamma)}(x) &=& \overline{D}(x)\Gamma U(x) \,, \\
{\cal O}_{\overline{U}U\pm\overline{D}D}^{(\Gamma)}(x)
    &=& \frac{1}{\sqrt{2}}\bigg(\overline{U}(x)\Gamma U(x)
        \pm \overline{D}(x)\Gamma D(x)\bigg) \,,
\end{eqnarray}
where $\Gamma$ denotes any product of Dirac matrices.
Di-quarks in the SU(2) color theory couple to the following local
operators
\begin{eqnarray}
{\cal O}_{UD}^{(\Gamma)}(x) &=& U^T(x)(-i\sigma^2)C\Gamma D(x) \,, \\
{\cal O}_{DU}^{(\Gamma)}(x) &=& D^T(x)(-i\sigma^2)C\Gamma U(x) \,, \\
{\cal O}_{UU\pm DD}^{(\Gamma)}(x)
    &=& \frac{1}{\sqrt{2}}\bigg(U^T(x)(-i\sigma^2)C\Gamma U(x)
        \pm D^T(x)(-i\sigma^2)C\Gamma D(x)\bigg) \,,
\end{eqnarray}
where the Pauli structure $-i\sigma^2$ acts on color indices while the
charge conjugation operator $C$ acts on Dirac indices. In \cite{Lewis:2011zb,Hietanen:2013fya} the meson masses were extracted from the two-point correlation functions
\begin{align}
C^{(\Gamma)}_{\overline{U}D}(t_i-t_f)
 & =  \sum_{\vec x_i,\vec x_f} \left\langle {\cal O}_{UD}^{(\Gamma)}(x_f)
{\cal O}_{UD}^{(\Gamma)\dagger}(x_i) \right\rangle.\nonumber\\
 & = \sum_{\vec x_i,\vec x_f} \Tr \Gamma
S_{D\overline{D}}(x_f,x_i)\gamma^0\Gamma^\dagger\gamma^0 S_{U\overline{U}}(x_i,x_f),
\end{align}
where $S_{U\overline{U}}(x,y) = \langle
U(x)\overline{U}(y)\rangle$. The quantities of interest are
pseudoscalar $\Gamma=\gamma_5$, vector $\Gamma=\gamma_k$ ($k=1,2,3$),
and axial vector $\Gamma=\gamma_5\gamma_k$ mesons. 

The second point to answer is: At what energy scale one can hope to discover new states such as spin-one resonances?  

Having both a well defined underlying gauge theory as well as its lattice simulations we can provide the first preliminary predictions \cite{Hietanen:2013fya}. In units of $2\sqrt{2}f \simeq (246/s_\theta)$~GeV we have
\begin{equation}
\frac{m_\rho}{2\sqrt{2} \, f} \simeq   10.20(0.16)(1.14)(1.22)\, \ , \qquad \frac{m_A}{2\sqrt{2}f} \simeq 13.20(0.53)(1.50)(1.50) \ , 
\end{equation}
where the first error is statistical, the second comes from the continuum extrapolation and the third from the uncertainty in $Z_a$ \cite{Hietanen:2013fya}. In electroweak physical units
\begin{equation}
{m_\rho} \simeq   2510(40)(280)(300)~{\rm GeV}/s_\theta\, \ , \qquad {m_A}\simeq 3270(130)(370)(370) ~{\rm GeV}/s_\theta\ . 
\end{equation}
The phenomenological lightest vector mesons occur for $\theta=\pi/2$ associated to the Technicolor direction while for small $\theta$, which is the direction associated to a prevalently pseudo-GB Higgs, the vectors become very heavy to be easily detectable even for the next generation of colliders.

One can imagine to use a more involved gauge theory dynamics by, for example, adding new type of matter singlet with respect to the electroweak interactions as advocated in \cite{Dietrich:2006cm} in order to reach near-conformal dynamics. An explicit construction involving directly the SU(2) gauge group which besides two Dirac fermions in the fundamental also features fermions in the adjoint representation was considered explicitly in \cite{Ryttov:2008xe}. These theories might feature parametrically lighter non-GB states because they might feel the presence of a nearby infrared fixed point\footnote{This behavior requires that, as function of the parameter space of the theory, such as the number of flavors, the transition to conformality is smooth, i.e. no jumping phenomenon exists \cite{Sannino:2012wy} in the form of a first-order phase transition.}. Lattice investigations of this type of dynamics are underway \cite{Fodor:2014pqa}.

 \section{Conclusions and Outlook}
\label{conclusions}

Models of composite (Goldstone) Higgs are relevant alternatives to supersymmetric extensions of the SM. 
In this paper we  have explicitly shown that by providing an explicit UV completion, even if only partial, for models of composite Higgs, based on a strongly interacting gauge theory with fermionic matter fields, offers a unified framework where one can study simultaneously models of pseudo-GB Higgses and Technicolor models. In the Technicolor limit the Higgs is identified with the lightest scalar resonance of the dynamics.
We can, therefore, answer several questions that cannot be otherwise addressed via the mere knowledge of the global symmetries of the effective theory such as: How heavy are the vector-mesons? Is it there a stable dark matter candidate? Does the global symmetry break according to the phenomenologically desired pattern?

We focused on the well known example of the flavor symmetry SU(4)/Sp(4)$\sim$SO(6)/SO(5) breaking pattern. The most minimal strongly coupled gauge theory able to achieve this breaking pattern is the SU(2) gauge theory with 4 Weyl fermions transforming according to the fundamental representation of the gauge group. Henceforth, seen from a fundamental dynamics point of view, this is also the minimal symmetry breaking pattern to serve as foundation for a pseudo-GB Higgs model as well as minimal models of Technicolor.
The coset space contains 5 pGBs, three of which are eaten by the massive $W$ and $Z$ assuming that a chiral condensate develops breaking spontaneously the electroweak symmetry. The fate of the remaining pGBs depends on the way the complete dynamics aligns the electroweak theory with respect to the vacuum condensate of the theory: in the Technicolor alignment, they form a complex stable dark matter candidate, while in the pGB Higgs alignment one state plays the role of the discovered Higgs boson and the other state remains neutral with respect to the electroweak symmetry but it is not expected to be stable. 
Our analysis shows that the Technicolor alignment is more natural as it is preferred by the top loop corrections, while to achieve the pGB Higgs vacuum one needs to introduce yet another explicit SU(4) symmetry breaking operator with an ad-hoc tuned coupling.

On general grounds, however, there is no reason to expect the condensate to align in either the pure Technicolor or pGB Higgs limit. We should rather conclude that, in the absence of a complete theory of SM mass generation a well as explicit breaking of the SU(4) symmetry, the vacuum alignment angle $\theta$ can assume any value between $0$ and $\pi/2$. For a generic alignment there is a relevant mixing between the pGB and the techni-Higgs. In this case neither the pure pGB nor the techni-Higgs state can be identified with the observed Higgs, but will be the lightest eigenstate. We have also demonstrated that this mixing affects the physical couplings of the new Higgs and argued that the additional singlet cannot play the role of a stable dark matter candidate unless the condensate is aligned mostly in the Technicolor vacuum.
Last but not the least, having a well defined UV completion allowed to use recent lattice computations to predict masses and couplings of heavy states, like for instance the lightest spin-one resonances (techni-$\rho$ and techni-axial), for a generic vacuum alignment. It turns out that the lightest state is the techni-$\rho$ but it is already rather heavy, with a mass above 2.5 TeV (the heavier the closer to the pGB Higgs alignment). We could therefore argue that, for the LHC, the scalar sector can be more directly tested while the spin-one states are rather challenging to explore. 

Our analysis can be applied to other patterns of symmetry breaking, like for instance SU(6)$\to$Sp(6), which is the minimal case with two pseudo-GB Higgs doublets.

We aimed at bridging the gap among different models of composite dynamics at the electroweak scale both at the effective model description as well as at a more fundamental level. Having at our disposal the fundamental description of, at least, one relevant piece of the composite gauge dynamics, we were able to make relevant physical predictions for the most elusive part of any low energy effective description, i.e. the mass scale of the new spin-one states to be discovered at colliders. 

Much remains to be investigated, at the fundamental level, from a model building point of view and last but not the least experimentally in order to rule-out models of composite dynamics at the electroweak scale.

\acknowledgments 
The work of F.S. is partially supported by the Danish National Research Foundation under the grant number DNRF:90. G.C. acknowledges partial support from 
the Labex-LIO (Lyon Institute of Origins) under grant ANR-10-LABX-66. G.C. would also like to thank the ESF Holograv Programme for supporting his participation to workshops in Edinburgh and Swansea, where the inspiration for this work was born.

\end{document}